\RequirePackage[2020-02-02]{latexrelease}
\documentclass[aps,prd,groupedaddress,graphicx,nofootinbib]{revtex4}
\usepackage{amsmath,amssymb,graphics,graphicx,color,epsf}
\usepackage{subfigure}
\usepackage[scanall]{psfrag}  % the scanall option enables \tex inside EPS files
\usepackage{tikz}

\begin{document}

\title{D-term inflation in braneworld models: Consistency with cosmic-string bounds and early-time Hubble tension resolving models}

\author{Chia-Min Lin}

\affiliation{Fundamental General Education Center, National Chin-Yi University of Technology, Taichung 41170, Taiwan}

%\baselineskip 14pt

%\date{Draft \today}

\begin{abstract}

We revisit D-term inflation with the bounds of the cosmic string tension from gravitational wave observations and consider the possible deviation of the spectral index compared with the $\Lambda$CDM model in light of prerecombination resolutions of the Hubble tension. D-term inflation requires very small coupling constants under these constraints. We show that natural coupling constants ($g=\lambda=0.1$) can be achieved in the case of D-term inflation on the brane. 

\end{abstract}
\maketitle
\large
\baselineskip 18pt
\section{Introduction}

Since the observation of gravitational waves \cite{TheLIGOScientific:2016wyq, Abbott:2017xzu}, we are entering a new era of cosmology. More precise measurements would be provided by future experiments such as LIGO \cite{TheLIGOScientific:2014jea}, Virgo \cite{TheVirgo:2014hva}, LISA \cite{Audley:2017drz}, DICIGO/BBO \cite{Yagi:2011wg}, the Einstein Telescope (ET) \cite{Punturo:2010zz}, Cosmic Explorer (CE) \cite{Evans:2016mbw}, Taiji \cite{Guo:2018npi}, and TianQin \cite{Mei:2020lrl}. This is bound to have profound ramifications on cosmology.

The idea of cosmic inflation \cite{Starobinsky:1980te, Sato:1980yn, Guth:1980zm, Linde:1981mu} solves many problems (such as the horizon, flatness, and unwanted relics problems\footnote{What kind of relics are unwanted depends on whether there are monopoles, gravitinos, or something else beyond the standard model of particle physics which may be overabundantly produced in a hot big bang.}) of old hot big bang model and explains why our universe is so big (compared with Planck scale), so long-lived (compared with Planck time), geometrically so flat (and not perfectly flat), and have a large entropy with a variety of things. It is intriguing that in this scenario, the origin of all the structures of the universe (such as galaxies and planets) is ultimately from quantum fluctuations. It is arguably the standard scenario of the very early universe. Yet we are still searching for the best inflation model.

Among the inflationary models, D-term inflation \cite{Binetruy:1996xj, Halyo:1996pp} is a supersymmetric (SUSY) realization of hybrid inflation  \cite{Linde:1993cn}. Hybrid inflation provides an effective way to produce small field inflation which is defined to be inflation models with an inflaton field value smaller than the Planck scale (at least when our observable universe is leaving the horizon). Although small field inflation models also produce primordial gravitational waves, the value of tensor-to-scalar ratio $r$ is typically too small to be observed in near-future experiments \cite{Lyth:1996im}. Instead, one can study a class of small-field inflation models in which cosmic strings are produced after inflation via gravitational waves generated by those cosmic strings \cite{Lin:2021wbn}. A salient feature of D-term inflation is the cosmic string production after inflation. The energy per unit length of cosmic strings (also known as the string tension $\mu$) can be significantly large and it can impose signatures for observations. String tension is commonly expressed via a dimensionless combination $G\mu$, where $G$ is Newton's constant. Current constraint from CMB (cosmic microwave background) measurement is $G\mu<1.1 \times 10^{-7}$ \cite{Charnock:2016nzm}. More precise measurements can be obtained through observations of gravitational waves since the vibration of cosmic strings generate gravitational radiation \cite{Hindmarsh:1994re}. Current limit from the European Pulsar Timing Array (EPTA) implies $G\mu \lesssim 10^{-11}$ \cite{vanHaasteren:2011ni}. Recently there is a possible signal of stochastic gravitational waves background from NANOGrav Collaboration \cite{Arzoumanian:2020vkk}. This corresponds to a string tension $G\mu \in (4\times 10^{-11}, 10^{-10})$ at the $68 \%$ confidence level \cite{Ellis:2020ena}\footnote{In \cite{Blasi:2020mfx}, a slightly different value of $G\mu \in (6\times 10^{-11}, 1.7\times 10^{-10})$ at the $68 \%$ confidence level is obtained. There are some uncertainties regarding the constraints due to the types and modeling of the string network. In some models, there is a stronger constraint of $G\mu \lesssim 4\times10^{-15}$  \cite{LIGOScientific:2021nrg}.}. In the calculation of the following sections, we will be using $G\mu=10^{-11}$.

On the other hand, recently there is a discrepancy between observations of the Hubble constant $H_0$ between high redshift (such as observations of CMB and baryon acoustic oscillations (BAO)) and low redshift measurements using local distance ladder (such as Cepheids and SNe Ia). The measurement from CMB is \cite{Planck:2018vyg}
\begin{equation}
H_0=67.4 \pm 0.5 \mbox{ km s}^{-1}\mbox{ Mpc}^{-1}.
\label{high}
\end{equation}
However, the measurement from the SH0ES collaboration is \cite{Riess:2021jrx}
\begin{equation}
H_0=73.04 \pm 1.04 \mbox{ km s}^{-1}\mbox{ Mpc}^{-1}.
\label{low}
\end{equation}
There are other experiments and the discrepancy cannot be easily explained by systematic error \cite{Efstathiou:2013via, Addison:2015wyg, Planck:2016tof, Aylor:2018drw, Verde:2019ivm}.
This discrepancy is known as the Hubble tension (see \cite{DiValentino:2021izs, Schoneberg:2021qvd, Abdalla:2022yfr, Perivolaropoulos:2021jda} for recent reviews) and currently the disagreement is about $4\sigma$ to $6\sigma$.
Since the result from CMB measurement is based on the $\Lambda$CDM model, many proposed resolutions for the Hubble tension assume some modifications of the $\Lambda$CDM model. In this case, there may be corresponding modifications of the spectral index \cite{DiValentino:2018zjj, Ye:2021nej, Jiang:2022uyg}. According to observations based on $\Lambda$CDM model, the spectral index is given by $n_s=0.965 \pm 0.004$ \cite{Planck:2018vyg}. On the other hand, in prerecombination resolutions of the Hubble tension (such as early dark energy), the Monte Carlo Markov chain analysis done by \cite{Ye:2021nej} shows 
\begin{equation}
\delta n_s \simeq 0.4 \frac{\delta H_0}{H_0},
\end{equation}
for lifting high redshift $H_0$ and it seems to be pointing to\footnote{Very roughly, from Eqs.~(\ref{high}) and (\ref{low}) we can calculate $n_s=0.96+0.4\frac{73.04-67.4}{70}=0.99 \sim 1$.} 
\begin{equation}
n_s=1. 
\label{n1}
\end{equation}
This increase of $n_s$ compared with  $\Lambda$CDM model was consistent with some earlier works \cite{Poulin:2018cxd, Agrawal:2019lmo, Lin:2019qug, Ye:2020oix, Ye:2020btb}, where the results indicate that in order to have $H_0 \gtrsim 71 \mbox{ km s}^{-1}\mbox{ Mpc}^{-1}$, the spectral index is 
\begin{equation}
n_s \gtrsim 0.98.
\label{n98}
\end{equation} 
The physical meaning of the increase of $n_s$ is to compensate for the suppression of small-scale fluctuations in those models.  
In \cite{Takahashi:2021bti}, cosmological implications of $n_s \sim 1$ are considered, in particular on the model of axion curvaton models.  
Here we focus on the implication of $n_s \sim 1$ to D-term inflation.

\section{D-term Inflation}
\label{sec2}
D-term inflation is a SUSY hybrid inflation. The superpotential is given by \cite{Binetruy:1996xj, Halyo:1996pp}
\begin{equation}
W_D=\lambda S \Phi_+ \Phi_-
\end{equation}
where $S$ is the inflaton superfield, $\lambda$ is the superpotential coupling and $\Phi_\pm$ are chiral superfields charged under the $U(1)_{FI}$ gauge symmetry responsible for the Fayet-Iliopoulos term. The corresponding SUSY tree-level effective scalar potential is
\begin{equation}
V(S, \Phi_+, \Phi_-)=\lambda^2 \left[ |S|^2\left( |\Phi_+|^2+|\Phi_-|^2 \right)+|\Phi_+|^2|\Phi_-|^2 \right]+\frac{g^2}{2}\left( |\Phi_+|^2-|\Phi_-|^2 +\xi \right)^2,
\end{equation}
where $\xi$ is the Fayet-Iliopoulos term and $g$ is the $U(1)_{FI}$ gauge coupling. From the potential, the true vacuum is given by
\begin{equation}
\langle S \rangle=0,\;\;\;\; \langle \Phi_+ \rangle =0,\;\;\;\; \langle \Phi_- \rangle =\sqrt{\xi}.
\end{equation}
During inflation, when the inflaton field value is larger than the critical value $|S| \gg |S|_c=g\xi^{1/2}/\lambda$ the field value of $S$ provides effective masses to $\Phi_+$ and $\Phi_-$, which drives their field values to zero. The potential minimum is along a flat valley and is given by
\begin{equation}
V=V_0=\frac{g^2\xi^2}{2}.
\label{eq7}
\end{equation}
The 1-loop inflaton potential is 
\begin{equation}
V(S)=V_0+\frac{g^4\xi^2}{32\pi^2}\left[2 \ln \left( \frac{\lambda^2 |S|^2}{M^2} \right) +(z+1)^2 \ln (1+z^{-1})+(z-1)^2 \ln (1-z^{-1}) \right],
\end{equation}
where $z=\lambda^2|S|^2/g^2\xi=|S|^2/|S|^2_c$ and $M$ is a renormalization scale. If $z \gg 1$, the 1-loop potential can be approximated by
\begin{equation}
V(s)=V_0+\frac{g^4 \xi^2}{16 \pi^2}\ln \left( \frac{s^2}{2M^2} \right),
\label{eq2}
\end{equation}
where $S \equiv s/\sqrt{2}$. In the following calculations, we assume $V_0$ dominates the potential, namely $V \simeq V_0$ as hybrid inflation should have.
The slow-roll parameters are given by
\begin{equation}
\epsilon\equiv \frac{M_P^2}{2}\left( \frac{V^\prime}{V} \right)^2=\frac{M_P^2 g^4}{32 \pi^4 s^2},
\label{epsilon}
\end{equation}
and 
\begin{equation}
\eta \equiv M_P^2\frac{V^{\prime\prime}}{V}=-\frac{M_P^2g^2}{4\pi^2s^2},
\label{eta}
\end{equation}
where $M_P=2.4 \times 10^{18}$ GeV is the reduced Planck mass.
The number of $e$-folds $N$ is given by
\begin{equation}
N=\frac{2\pi^2}{g^2}\left(s^2-s_e^2 \right),
\label{efolds}
\end{equation}
where $s_e$ marks the end of inflation. Inflation ends either when the inflaton drops below its critical value $s_c=g\sqrt{2}\sqrt{\xi}/\lambda$ or when the second slow-roll parameter becomes $|\eta|=1$ at $s_{s.r.}=(g/2\pi)M_P$ as can be seen from Eq.~(\ref{eta}). Namely,
\begin{equation}
s_e=\mbox{max}(s_{s.r.},s_c).
\end{equation} 
Depending on the mechanism of reheating, $N$ is roughly $50 \lesssim N \lesssim 60$. We will take $N=60$ when a numerical calculation is needed in the following.
It will be useful to calculate
\begin{equation}
\frac{s_c^2}{s_{s.r.}^2}=\frac{8\pi^2 \xi}{\lambda^2 M_P^2}.
\label{ss}
\end{equation}
When $\lambda$ is small, we may have $s_c>s_{s.r.}$. In this case, from Eqs.~(\ref{eta}), (\ref{efolds}), and (\ref{ss}),
\begin{equation}
\eta=-\frac{1}{\frac{s_c^2}{s_{s.r.}^2}+2N}.
\label{etasc}
\end{equation}
It is a novel feature of our calculation to express the results in terms of $s_c^2/s_{s.r.}^2$.
On the other hand, if $\lambda$ is big, we may have $s_c<s_{s.r.}$. In this case,
\begin{equation}
\eta=-\frac{1}{1+2N} \simeq -\frac{1}{2N}.
\label{etasr}
\end{equation}
Let us make a mnemonic rule here, when $s_c<s_{s.r.}$, we just set $s_c^2/s_{s.r.}^2=0$ so that Eq.~(\ref{etasc}) includes Eq.~(\ref{etasr}) and there is no need to duplicate equations in the following discussion\footnote{Practically there is no difference between $1+2N$ and $2N$. This is not only because $121 \simeq 120$ but we can have chosen say, $N=59.5$ from the beginning instead of $N=60$, and this choice is equally good concerning the uncertainty of $N$.}. 
From Eqs.~(\ref{epsilon}) and (\ref{eta}), we can see that 
\begin{equation}
|\epsilon|=\frac{g^2}{8\pi^2}|\eta|.
\label{smalle}
\end{equation} 
Therefore, we neglect $\epsilon$ compared with $\eta$ as we always consider $g \leq 0.1$. In particular, we have the spectral index given by
\begin{equation}
n_s = 1+ 2\eta-6\epsilon \simeq 1+2 \eta=1-\frac{2}{\frac{s_c^2}{s_{s.r.}^2}+2N}.
\label{index}
\end{equation}
The spectrum is 
\begin{equation}
P_R=\frac{1}{12\pi^2M_P^6}\frac{V^3}{V^{\prime 2}}=\frac{\xi^2}{6M_P^4}\left(\frac{s_c^2}{s_{s.r.}^2}+2N \right),
\label{spectrum}
\end{equation}
with CMB normalization given by $P_R^{1/2}=5 \times 10^{-5}$.
After inflation, the $U(1)$ gauge symmetry is spontaneously broken and cosmic strings form. The string tension $\mu$ (mass per unit length) is 
\begin{equation}
\mu=2\pi \xi.
\end{equation}
Experimental constraints for cosmic strings would give a constraint to $G\mu$ which by using the reduced Planck mass $M_P$ can be written as
\begin{equation}
G\mu=\frac{2\pi}{8\pi}\frac{\xi}{M_P^2}=\frac{\xi}{4 M_P^2}.
\label{string}
\end{equation}
Note that $\xi$ determines the scale of inflation via Eq.~(\ref{eq7}).

\section{Conventional D-term Inflation}

The coupling constants $g$ and $\lambda$ are free parameters in D-term inflation. However, aesthetically we may start from $g = \lambda = 0.1$. We refer to this as conventional D-term inflation. In this case, we have $s_c<s_{s.r.}$ due to the small value of $\xi$ and big value of $g$. This statement will be verified soon in the following calculation. From Eq.~(\ref{spectrum}) (and the mnemonic rule), we have
\begin{equation}
P_R=\frac{N\xi^2}{3 M_P^4}=\frac{20 \xi^2}{M_P^4}=(5 \times 10^{-5})^2
\end{equation}
This gives $\xi=1.1 \times 10^{-5} M_P^2$. 
By using Eq.~(\ref{ss}), we can now calculate
\begin{equation}
\frac{s^2_c}{s^2_{s.r.}}=8.8 \times 10^{-2} 
\end{equation}
to verify our previous assumption of $s_c<s_{s.r.}$.
From Eq.~(\ref{index}) (and the mnemonic rule), we obtain 
\begin{equation}
n_s=1-\frac{1}{N}=0.98,
\label{conven}
\end{equation}
which satisfies Eq.~(\ref{n98}).
The running spectral index can be obtained from $n_s$ as
\begin{equation}
\alpha=-\frac{dn_s}{dN}=-\frac{1}{N^2}=-0.00028.
\end{equation}
This is compatible with the Planck data $|\alpha|<0.01$ \cite{Planck:2018jri}.
From Eq.~(\ref{string}), we have $G\mu=2.8 \times 10^{-6}$. However, as discussed in the Introduction section, we need $G\mu=10^{-11}$ to achieve the current experimental bound.
Therefore, conventional D-term inflation is ruled out by experimental searches of cosmic strings. We refer to this as the cosmic string problem.

\section{D-term Inflation with small coupling constants}

It is shown in \cite{Endo:2003fr, Rocher:2004my} that cosmic string problem with constraints from CMB can be evaded if we relax the requirement of $g = \lambda = 0.1$. In \cite{Endo:2003fr}, the authors obtain $\lambda \lesssim O(10^{-4}-10^{-5})$. In \cite{Rocher:2004my}, the authors obtain $g \lesssim 2 \times 10^{-2}$ and $\lambda \lesssim 3 \times 10^{-5}$. In this section, we calculate the coupling constants $g$ and $\lambda$ by using the much more stringent constraint for cosmic strings from gravitational waves. If $G\mu=10^{-11}$, from Eq.~(\ref{string}) we have $\xi=4 \times 10^{-11}M_P^2$ or $\sqrt{\xi}=6.3 \times 10^{-6}M_P$.
Contrary to conventional D-term inflation, we have $s_c^2/s_{s.r.}^2 \gg 2N $ due to the smallness of $\lambda$ which will be verified later. From Eqs.~(\ref{ss}) and (\ref{spectrum}), we have
\begin{equation}
P_R=\frac{4\pi^2 \xi^3}{3M_P^6 \lambda^2}=25 \times 10^{-10}.
\label{e17}
\end{equation}
This implies
\begin{equation}
\lambda=1.8 \times 10^{-11},
\label{eq23}
\end{equation}
which is much smaller than those obtained in \cite{Endo:2003fr, Rocher:2004my} because we have updated the experimental constraint. In this case, from Eq.~(\ref{ss}), we have 
\begin{equation}
\frac{s_c^2}{s_{s.r.}^2}=9.4 \times 10^{12} \gg 2N=120,
\label{e19}
\end{equation}
which is in accordance with our assumption. For the allowed values of the coupling constant $g$, since we are considering a small-field inflation model, it is required that
\begin{equation}
s_c=\frac{g\sqrt{2}\sqrt{\xi}}{\lambda} < 0.1M_P.
\end{equation}
This implies
\begin{equation}
g<2.0 \times 10^{-7}.
\label{eq26}
\end{equation}
In addition, by noticing that Eq.~(\ref{e17}) is independent of $N$, we can conclude that the spectrum is scale-invariant, namely the spectral index $n_s=1$ which satisfies Eq.~(\ref{n1})!\footnote{With $|n_s-1|$ smaller than $10^{-12}$, which can be seen from Eq.~(\ref{index}) and Eq.~(\ref{e19}).}. The requirement of cosmic string constraint drives the spectrum to be scale-invariant. Interestingly, this is in accordance with the proposals to alleviate the Hubble tension mentioned in the Introduction section. 

One may not be satisfied with the small coupling constants given in Eqs.~(\ref{eq23}) and (\ref{eq26}). Especially a very small gauge coupling $g$ seems to have difficulty connecting to known gauge couplings in (known) particle physics. In the following section, we propose a model to make the coupling constants bigger.

\section{D-term Inflation on the brane}

If our four-dimensional world is a 3-brane embedded in a higher-dimensional bulk, the Friedmann equation becomes \cite{Cline:1999ts, Csaki:1999jh, Binetruy:1999ut, Binetruy:1999hy, Freese:2002sq, Freese:2002gv, Maartens:1999hf}
\begin{equation}
H^2=\frac{1}{3M_P}\rho \left[ 1+ \frac{\rho}{2 \Lambda} \right],
\end{equation}
where $\Lambda$ provides a relation between the four-dimensional Planck scale $M_4=\sqrt{8 \pi}M_P$ and five-dimensional Planck scale $M_5$ via
\begin{equation}
M_4=\sqrt{\frac{3}{4\pi}}\left( \frac{M_5^2}{\sqrt{\Lambda}} \right)M_5.
\end{equation}
Here we consider $M_5$ as a free parameter that can be considerably smaller than $M_4$.
The nucleosynthesis limit implies that $\Lambda \gtrsim (1\mbox{ MeV})^4 \sim (10^{-21}M_P)^4$.
Stronger constraints up to $\Lambda \gtrsim 10^5\mbox{ MeV}^4$ can be obtained from Solar System tests \cite{Deng:2016ztc, Casadio:2015jva, Boehmer:2008zh}.
A more stringent constraint, $M_5 \gtrsim 10^5$ TeV can be obtained by requiring the theory to reduce to Newtonian gravity on scales larger than $1$ mm, this corresponds to 
\begin{equation}
\Lambda \gtrsim 5.0 \times 10^{-53} M_P^4.
\label{ls} 
\end{equation}
For D-term inflation on the brane, the slow-roll parameters\footnote{We use the same notations for the slow-roll parameters as in the previous sections, but confusion should not be caused due to the context.} are given by \cite{Maartens:1999hf, Lee:2009mj}
\begin{equation}
\epsilon=\frac{M_P^2}{2}\left( \frac{V^\prime}{V} \right)^2\frac{1}{\left( 1+\frac{V}{2\Lambda} \right)^2}\left( 1+\frac{V}{\Lambda} \right),
\end{equation}
and 
\begin{equation}
\eta=M_P^2 \frac{V^{\prime\prime}}{V} \left( \frac{1}{1+\frac{V}{2\Lambda}} \right).
\end{equation}
In this case, instead of Eqs.~(\ref{epsilon}) and (\ref{eta}), we have
\begin{equation}
\epsilon=\frac{M_P^2 g^4}{32 \pi^4 s^2}\frac{1}{\left( 1+\frac{g^2 \xi^2}{4\Lambda} \right)^2}\left( 1+\frac{g^2 \xi^2}{2\Lambda} \right),
\label{e36}
\end{equation}
and 
\begin{equation}
\eta=-\frac{M_P^2 g^2}{4\pi^2 s^2}\frac{1}{\left( 1+\frac{g^2 \xi^2}{4\Lambda} \right)}.
\label{eq25}
\end{equation}
We will consider the case $g^2\xi^2/2 \gg \Lambda$, therefore from Eqs.~(\ref{e36}) and (\ref{eq25}), we have
\begin{equation} 
|\epsilon|  \simeq  \frac{g^2}{4\pi^2}|\eta|,
\end{equation}
which can be compared with Eq.~(\ref{smalle}) therefore $\epsilon$ will be neglected again in the following discussion. 
The inflaton field value $s_{s.r.}$ when slow-roll fails can be obtained by solving $|\eta|=1$ as
\begin{equation}
s^2_{s.r.}=\frac{M_P^2 g^2}{4\pi^2 \left( 1+\frac{g^2 \xi^2}{4\Lambda} \right)}.
\label{bsr}
\end{equation}
The number of $e$-folds is 
\begin{equation}
N=\frac{2\pi^2}{g^2}\left( 1+\frac{g^2\xi^2}{4 \Lambda} \right) \left(s^2-s_e^2 \right),
\label{bn}
\end{equation}
where the end of inflation is determined by the inflaton field value
\begin{equation}
s_e=\mbox{max}(s_{s.r.},s_c).
\end{equation}
By using $s_c=g\sqrt{2}\sqrt{\xi}/\lambda$, we obtain
\begin{equation}
\frac{s_c^2}{s_{s.r.}^2}=\frac{8 \pi^2 \xi}{\lambda^2 M_P^2} \left( 1+\frac{g^2 \xi^2}{4 \Lambda} \right).
\label{bss}
\end{equation}
This can be compared with Eq.~(\ref{ss}).
When $s_c>s_{s.r.}$, from Eqs.~(\ref{eq25}), (\ref{bn}), and (\ref{bss}), $\eta$ can be written as
\begin{equation}
\eta=-\frac{1}{\frac{s_c^2}{s_{s.r.}^2}+2N}.
\label{eq27}
\end{equation} 
The spectral index is given by
\begin{equation}
n_s = 1+ 2\eta-6\epsilon \simeq 1+2 \eta=1-\frac{2}{\frac{s_c^2}{s_{s.r.}^2}+2N}.
\label{bindex}
\end{equation}
Note that Eqs.~(\ref{eq27}) and (\ref{bindex}) appears to have the same form as Eq.~(\ref{etasc}) and (\ref{index}), but the corresponding $s_{s.r.}$ are different.
When $s_c<s_{s.r.}$, our previous rule to neglect the factor $s_c^2/s_{s.r.}^2$ still applies.
The spectrum is 
\begin{equation}
P_R=\frac{1}{12\pi^2M_P^6}\frac{V^3}{V^{\prime 2}}\left( 1+ \frac{V}{2\Lambda} \right)^3=\frac{\xi^2}{6M_P^4}\left( 1+\frac{g^2\xi^2}{4 \Lambda} \right)^2 \left(\frac{s_c^2}{s_{s.r.}^2}+2N \right)
\label{bspectrum}
\end{equation}
In the following, we define
\begin{equation}
\left( 1+\frac{g^2\xi^2}{4 \Lambda} \right) \equiv L,
\end{equation}
since this factor appears a lot.
Our purpose here is to avoid small couplings, therefore we start by assuming $g=\lambda=0.1$. In this case, $s_c<s_{s.r.}$ as will be verified later. From Eq.~(\ref{bspectrum}) (and setting $s_c^2/s_{s.r.}^2=0$), the spectrum is
\begin{equation}
P_R=\frac{\xi^2L^2N}{3M_P^4}=25 \times 10^{-10}.
\end{equation}
By using $N=60$ and $\xi=4 \times 10^{-11}M_P^2$, we obtain $L=2.8 \times 10^5$. This can be achieved if $\Lambda=1.4 \times 10^{-29}M_P^4$, which satisfies Eq.~(\ref{ls}).
We can now calculate (by using Eq.~(\ref{bss}))
\begin{equation}
\frac{s^2_c}{s^2_{s.r.}}=\frac{8\pi^2 \xi L}{\lambda^2 M_P^2}=8.8 \times 10^{-2} 
\end{equation}
to verify our previous assumption of $s_c<s_{s.r.}$.

From Eq.~(\ref{eq25}), one may naively guess that $\eta$ would be very small since we have $L \sim O(10^5)$ in the denominator. However, from Eq.~(\ref{eq27}) (and our mnemonic rule of setting $s_c^2/s_{s.r.}^2=0$ when $s_c<s_{s.r.}$), we actually have $\eta=-1/2N$ and this implies $n_s=1+2\eta=0.98$. This spectral index is the same as Eq.~(\ref{conven}) of conventional D-term inflation, which satisfies Eq.~(\ref{n98}). Here we have assumed $g=\lambda=0.1$, but if we allow $\lambda$ to be smaller, it is possible to obtain the scale-invariant spectrum $n_s =1$. In order to show it, firstly, we assume $g=0.1$, $\lambda=10^{-3}$ and $s_c^2/s_{s.r.}^2 \gg 2N$. The spectrum is then
\begin{equation}
P_R=\frac{4\pi^2 \xi^3 L^3}{3M_P^6 \lambda^2}=25 \times 10^{-10}.
\label{eq51}
\end{equation} 
By using $\xi=4 \times 10^{-11}M_P^2$, we obtain $L=1.4 \times 10^5$ (which corresponds to $\Lambda=2.9 \times 10^{-29}M_P^4$). From Eq.~(\ref{bss}),
\begin{equation}
\frac{s^2_c}{s^2_{s.r.}}=\frac{8\pi^2 \xi L}{\lambda^2 M_P^2}=442,
\end{equation}
which is larger (although not too much larger) than $2N=120$. Secondly, we can consider $g=0.1$ and $\lambda=10^{-4}$. Through similar calculations, we have $L=3.1 \times 10^4$ (which corresponds to $\Lambda=1.3 \times 10^{-28}M_P^4$ ) and $s_c^2/s_{s.r.}^2=9790$. Note that for $\lambda \sim O(10^{-3}-10^{-4})$ the requirement of small field inflation $s_c=g\sqrt{2}\sqrt{\xi}/\lambda < 0.1M_P$ is satisfied even with $g$ as large as $g=0.1$. From Eq.~(\ref{bindex}), we can see that $n_s$ is driven to $n_s=1$ (which satisfied Eq.~(\ref{n1})) by having a smaller $\lambda$. Comparing Eq.~(\ref{eq51}) with Eq.~(\ref{e17}), we can understand the reason why $\lambda$ is not as small as the case in the previous section. It is because the condition of small $\lambda^2$ is now changed to small $\lambda^2/L^3$. Thanks to large $L$, $\lambda$ need not be so small.

\section{Conclusion and Discussion}
\label{con}

The simplest D-term inflation can be consistent with the cosmic string bound provided by observations of gravitational waves with very small coupling constants. This drives the spectral index to $n_s=1$ which may be an interesting result in light of the Hubble tension. We show that in the case of D-term inflation on the brane, the coupling constants can be $g=\lambda=0.1$. In this case, we have $n_s=0.98$. If we lower one coupling constant to $\lambda<10^{-3}$, the spectral index $n_s=1$ can be achieved.

To some extent, the requirement of small coupling constants for the simplest model to work is intuitively expected. From $s_{s.r.}=(g/2\pi) M_P$ we can see that although D-term inflation is a small field inflation model, the field value is not far below $M_P$ for a large $g$. This means the energy scale of the potential is about the scale of grand unified theories (GUT). Therefore, the tension of cosmic strings would also be large. On the other hand, D-term inflation on the brane can effectively reduce the inflation scale via the $L$ factor. Future constraints on extra dimensions would provide a better understanding of $M_5$, $\Lambda$, and the $L$ factor.

\acknowledgments
This work is supported by the National Science and Technology Council (NSTC) of Taiwan under Grant No. NSTC 111-2112-M-167-002.

\end{document}